\documentstyle[aps,prl,amsfonts,amssymb,floats,twocolumn,epsfig]{revtex}
\begin{document}
\draft
\twocolumn[\hsize\textwidth\columnwidth\hsize\csname
@twocolumnfalse\endcsname
\title{Observation of the Transverse Optical Plasmon in
SmLa$_{0.8}$Sr$_{0.2}$CuO$_{4-\delta}$}
\author{Diana Duli\'c$^{1}$, A. Pimenov$^{2}$, D. van der Marel$^{1}$,
D. M. Broun$^{3,4}$,  Saeid Kamal$^{3}$, W. N. Hardy$^{3}$, A. A.
Tsvetkov$^{1}$, I.M. Sutjaha$^{5}$, Ruixing Liang$^{3}$, A. A.
Menovsky$^{5}$, A. Loidl$^{2}$ and S. S. Saxena$^{1}$}
\address{$^{1}$Laboratory of Solid State Physics, Materials Science Centre,
Nijenborgh 4, 9747 AG Groningen, The Netherlands}
\address{$^{2}$Experimentalphysik V, University of Augsburg, Universitaetsstr. 2,
86135 Augsburg, Germany}
\address{$^{3}$Department of Physics and Astronomy, University of British
Columbia, Vancouver, BC, V6T 1Z1, Canada}
\address{$^{4}$Interdisciplinary Research Centre in Superconductivity and Department
of Physics, Cavendish Laboratory, University of Cambridge,
Madingley Road, Cambridge CB3 0HE, United Kingdom}
\address{$^{5}$Van der Waals - Zeeman Institut, University of Amsterdam, The Netherlands}
\date{oktober 25, 2000}
\maketitle
\begin{abstract}
We present microwave and infrared measurements on
SmLa$_{0.8}$Sr$_{0.2}$CuO$_{4-\delta}$, which are direct evidence
for the existence of a transverse optical plasma mode, observed as
a peak in the $c$-axis optical conductivity. This mode appears as
a consequence of the existence of two different intrinsic
Josephson couplings between the CuO$_2$ layers, one with a
Sm$_2$O$_2$ block layer, and the other one with a
(La,Sr)$_2$O$_{2-\delta}$ block layer. From the frequencies and the
intensities of the collective modes we determine the value of the 
compressibility of the two dimensional electron fluid in the 
copper oxygen planes.
\end{abstract}
\pacs{74.72.-h,74.25.Gz}
]
\narrowtext

In 1966 A.J. Leggett predicted for superconductors with two bands of
charge carriers a collective oscillation corresponding 
to small fluctuations of the relative phases of the two condensates, 
briefly indicated as excitons below the superconducting gap\cite{tony}. 
In principle these
excitons should be observable with electromagnetic radiation, but in practice
the effect on the infrared optical properties of most superconducting 
materials has been too small to be observable, except for, as we will demonstrate 
in the present paper, the bi-layer cuprate superconductors.
The cuprate high temperature superconductors naturally form weakly
coupled stacks of superconducting layers\cite{lawrence}. Some members of 
this family, {\em e.g.} Bi$_2$Sr$_2$CaCu$_2$O$_8$, have two superconducting 
layers per unit cell. These materials are realizations of a two-band 
superconductor, satisfying the following unique conditions:
(i) For polarization of the electric field perpendicular to the conducting 
planes the metallic screening is very weak due to strong anisotropy of the 
static and dynamical electrical conductivity. 
(ii) The dipole selection rules allow optical 
transitions which resonate at the Josephson exciton energy. 
 
\begin{figure}
 \centerline{\epsfig{figure=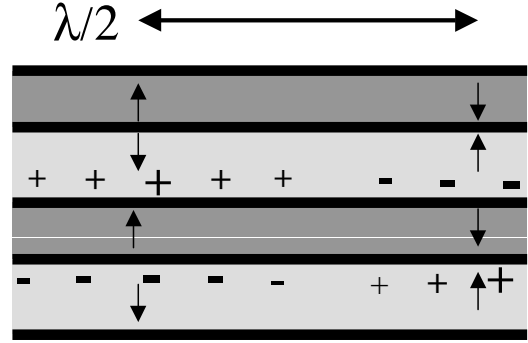,width=7cm,clip=}}
 \caption{Snapshot of the currents (arrows) and charges ($\pm$ signs) of the
 transverse {\em optical} JPR. This mode carries a 
 finite net polarization transverse to the direction of propagation.}
 \label{tojpr}
\end{figure}

In Ref.~\cite{lt21} two of us (DvdM and AAT)
calculated the dielectric function for cuprate
superconductors with two CuO$_2$ planes per unit cell, using
the Lawrence-Doniach model\cite{lawrence} with alternating coupling
constants (the 'multi-layer model'). A direct consequence was the 
presence of a transverse optical plasma mode, polarized perpendicular to the
planes for propagation along the planes. Similar to a
transverse optical phonon, and in contrast to the transverse 
Josephson plasma resonance (JPR)
in single layer cuprates, this mode is observable as an optically
allowed absorption in measurements of the optical conductivity. In
Fig.~\ref{tojpr} the currents and charge oscillations of these
modes are sketched. Recently the unusual 'bump' in the c-axis conductivity 
of YBCO\cite{markus,munzar} and Bi2212\cite{zelezny},
has been attributed to such a mode. However, due to the close proximity 
to the optical phonons it was not possible to separate these 
electronic collective modes from the optical lattice vibrations\cite{munzar}, 
which complicates the quantitative analysis of this interesting phenomenon.

Both in the two-band exciton model and in the multilayer model the collective modes
are oscillations of the relative phase of the two condensates, the inertia of which 
is due to the finite Josephson coupling between the two condensates. 
In Ref.~\cite{tony} the restoring force was provided by
the fact, that if $\delta N$ electrons are added to
a plane, the free energy increases with an amount $\delta F = \mu
\delta N + \delta N^2/(2Kn^2)$. Here $\mu$, $K$, and $n$ are the
chemical potential, the electronic compressibility, and the
electron density respectively. For a Fermi-liquid $Kn^2=\partial
N/ \partial \mu$ corresponds to density of states at the Fermi
level. In neutral fluids the compressibility causes
propagation of sound, whereas for electrons it causes the
dispersion of plasmons. 
On the other hand, in Ref.~\cite{lt21} the restoring force is 
provided by the long range Coulomb  interaction between each set 
of planes. Recently this model was extended by taking into account 
the compressibility \cite{lt21app}. Dispersion of the JPR in the 
cuprates due to the compressibility term has been previously described 
by Koyama and Tachiki\cite{koyama}. 

SmLa$_{0.8}$Sr$_{0.2}$CuO$_{4-\delta}$ forms the $T^*$
structure, with superconducting planes alternatingly separated by
two types of blocking layers, a
fluorite-type layer of Sm$_2$O$_2$ ($T'$ type) and a rocksalt-type
(La,Sr)$_2$O$_{2-\delta}$ ($T$ type) block layer. The lattice
parameter is 12.57 $\AA$, and the distances between CuO$_2$ layers
across the SmO and the LaO barriers are 6.13 $\AA$, and 6.43 $\AA$
respectively\cite{tokura}. Because the coupling across both
blocking layers is weak, the Josephson plasma resonances are well
below the optical phonon frequencies, allowing a clean separation
between the lattice vibrations and the electronic modes.
Recently {\em two} longitudinal Josephson plasma resonances have
been observed\cite{shibata} in embedded powder samples of this
material using the far-infrared sphere resonance method, 
but the transverse optical plasma mode could not be detected. 

Here we present infrared optical spectra of single crystal 
SmLa$_{0.8}$Sr$_{0.2}$CuO$_{4-\delta}$. We observed the
transverse optical plasma mode and the two longitudinal Josephson plasma resonances. 
The influence of the compressibility on the intensity and the frequency of the
observed resonances is large, a characteristic which these collective modes
have in common with the two-band excitons predicted in 1966\cite{tony}. Using a 
quantitative analysis of the intensities and peak positions of the three collective 
modes we extract the values of the intrinsic Josphson coupling constants
and the compressibility of the two dimensional charge fluid in the planes.

The single crystals of SmLa$_{0.8}$Sr$_{0.2}$CuO$_{4-\delta}$ were
grown at the University of Amsterdam using a four mirror image
furnace. They were annealed at 900 $^{\circ}$C for 24 hours to
homogenize the strontium concentration. The crystal was cut into
several slabs for the various experiments, and then annealed in a
high pressure oxygen furnace at the University of British
Columbia. The samples were slow cooled (one $^{\circ}$C per hour)
from 500$^{\circ}$C to 350$^{\circ}$C at 80 bars oxygen pressure,
held at 350$^{\circ}$C for 48 hours and then quenched by opening
the furnace. The large crystal faces (containing the $c$-axis)
were polished to an optical finish using diamond polishing pads.
The magnetization, as measured in a field of $2$ Oe with $H$
parallel to $c$, gave a midpoint $T_{c}$ of about 16K with a
transition width of $\pm$2K.

Three types of experiments were performed: (a) polarized
reflection measurements using a Fourier transform spectrometer in
the frequency range between 12 and 1000 cm$^{-1}$, and the
temperature range from 4 to 300K; {\em in-situ} evaporated gold
films were used as a reference; (b) polarized transmission
measurements in the frequency range from 4 to 20 cm$^{-1}$ using a
Mach-Zehnder interferometer arrangement\cite{andrei}, which allows
measurements of both transmission and phase shift (from the
Fresnel formulae, the absolute values of the complex dielectric
function can be determined directly from the measurements); and
(c) microwave cavity perturbation measurements at 1, 25 and 39
GHz. For (a), the dimensions of the sample were
0.7mm$\times$5mm$\times$6mm for the $a$, $b$ and $c$ directions
respectively. For (b), the $a$-dimension was 110$\mu$m, and for
(c) the $a$-dimensions were between 38 and 56$\mu$m.

\begin{figure}
 \centerline{\epsfig{figure=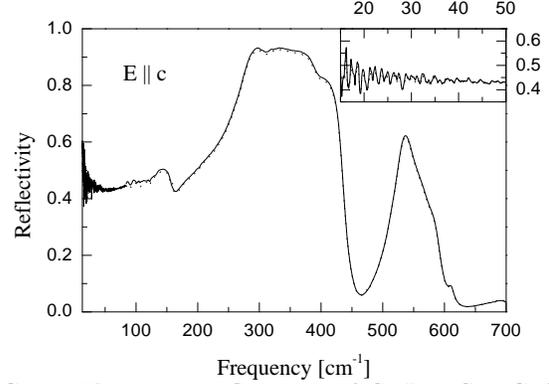,width=7cm,clip=}}
 \caption{The $c$-axis reflectivity of
 SmLa$_{0.8}$Sr$_{0.2}$CuO$_{4-\delta}$: 4K-solid line, 30K-dotted
 line. Inset: blown up region from 11 to 50cm.$^{-1}$}
 \label{reflect}
\end{figure}

\begin{figure}[t]
 \centerline{\epsfig{figure=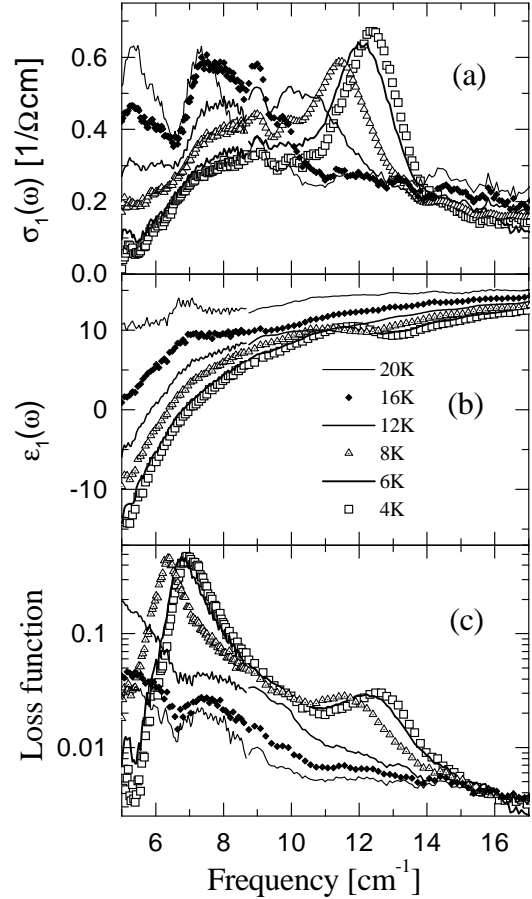,width=7cm,clip=}} 
 \caption{(a) Real part of the $c$-axis optical conductivity 
 for 4, 6, 8, 10 16 and 20 K. 
 (b) Real part of the $c$-axis dielectric function. The temperature labels 
 corresponding to all three panels are indicated.
 (c) The c-axis loss function, Im$\epsilon(\omega)^{-1}$. }
 \label{everig}
\end{figure}

In Fig.~\ref{reflect} we show the $c$-axis reflectivity spectra of
SmLa$_{0.8}$Sr$_{0.2}$CuO$_{4-\delta}$ above and below T$_{c}$, in
the frequency range from 13-700 cm $^{-1}$. In the inset of
Fig.~\ref{reflect} the frequency region from 13-50 cm$^{-1}$ has
been shown in more detail. In this region the sample has become
relatively transparent, and one can see interferences from the
copper block at the back side of the sample. These Fabry-Perot
interferences are absent from the mm-wave experiments, due to the
fact that we used much thinner samples in that case. The
important conclusion is that there is almost no temperature
dependence of the spectra in the 13-700 cm$^{-1}$ frequency
region. We measured transmission and phase shift data in the
frequency region from 4-20 cm$^{-1}$, for various temperatures.
The corresponding optical properties are obtained directly by the
use of the Fresnel formulae and the relation between the
dielectric function and optical conductivity: $\epsilon=4\pi
i\sigma/\omega$. The real part of the optical conductivity, the
real part of the dielectric function and the loss function
$Im(-1/\epsilon)$ are shown in Fig.~\ref{everig} (a), (b), and (c),
respectively.

In Ref.~\cite{lt21} the dielectric function has been derived for a
material with two superconducting layers per unit cell. The
crystal was treated as a stack of superconducting layers of
infinitesimal thickness, with two different Josephson coupling
constants $J$ and $K$, distances between the
neighbouring planes $x_I d$ and $x_K d$, $x_I+x_K=1$, and $d$ is
the lattice constant along the c-direction. 
The total dielectric function was shown to be of the form
\begin{equation}
  \frac{\epsilon_{av}^2}{\epsilon(\omega)} = 
  \frac{\tilde{z}_I\omega^2}{\omega(\omega+i\gamma_{I})-\omega_{I}^2 }
  +
  \frac{\tilde{z}_K\omega^2}{\omega(\omega+i\gamma_{K})-\omega_{K}^2 }
 \label{eq:difun2},
\end{equation}
Here the average background dielectric constant is
$\epsilon^s_{av} = (x_I/\epsilon^s_I + x_K/\epsilon^s_K)^{-1}$,
and $\epsilon^s_{\mu}$, $\omega_{\mu}$, and $\gamma_{\mu}$
are the local background dielectric constants
the screened Josephson plasma frequencies and the quasi-particle damping
characterizing the two types of junctions. 
Re$\epsilon(\omega)$ has two zero crossings corresponding
to the two longitudinal plasmons, and a pole at 
the {\em transverse} optical plasma frequency. The transverse
and the longitudinal plasma resonances show up as a peaks in the optical 
conductivity and the loss function respectively.
The transverse optical plasma frequency
$\omega_T = (\tilde{z}_I \omega_K^2 + \tilde{z}_K \omega_I^2)^{1/2}$,
and its inverse lifetime
$\gamma_T = \tilde{z}_I \gamma_K + \tilde{z}_K \gamma_I$
are fixed by the values of $\omega_I$, $\omega_K$, $\gamma_I$, $\gamma_K$,
and the weight factors
$\tilde{z}_K = 1-\tilde{z}_I$, which are a function of the 
relative volume fractions, the local dielectric functions, and the 
compressibility which we will discuss below.

\begin{figure}[t]
 \centerline{\epsfig{figure=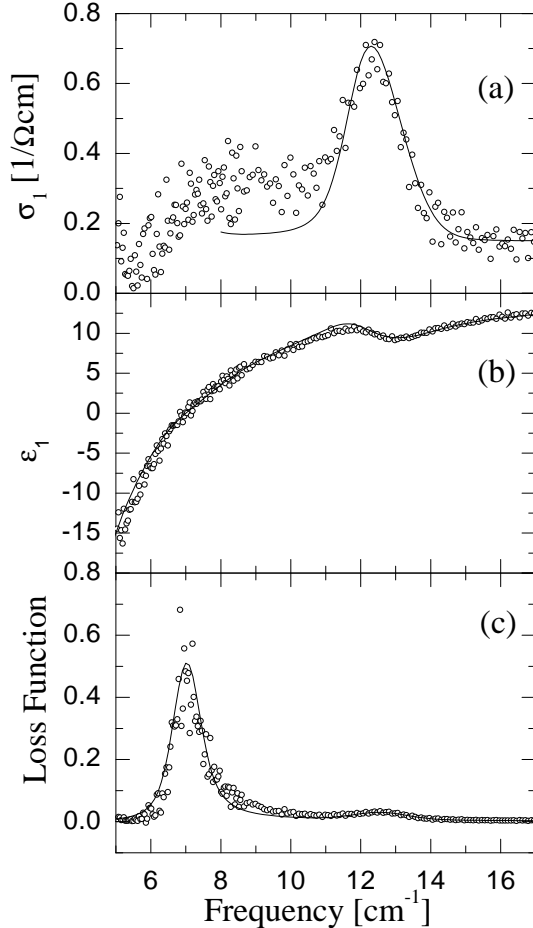,width=7cm,clip=}} \caption{(a)
 Fit (solid line) to (a) the real part of the $c$-axis optical
 conductivity, (b) Real part of the $c$-axis dielectric function, 
 (c) the loss function at 3K. The open circles are the experimental data.}
 \label{fit}
\end{figure}

The two longitudinal plasma frequencies can be read directly from the 
peak positions in the loss function, providing $\omega_I/2\pi c=7$cm$^{-1}$ 
and $\omega_K/2\pi c=12.6$cm$^{-1}$. 
The background dielectric constant is obtained by analyzing 
the reflectivity spectra of Fig.~\ref{reflect},
providing $\epsilon^s_{av} = 23.0 \pm 0.5$. The intensities of the two
peaks in the loss function, or, more accurately, the intensity of
the peak in $\sigma$ at $\omega_T$, provide the weight factor
$\tilde{z}_K$: $\int\sigma_{peak}(\omega)d\omega=
\tilde{z}_K\tilde{z}_I\epsilon^s_{av}(\omega_K^2-\omega_I^2)/\omega_T^{2}$.
This way we obtain $\tilde{z}_K = 0.04 \pm 0.005$, and $\tilde{z}_I = 0.96 \pm 0.005$.
In Fig.~\ref{fit} we show the theoretical spectra of the conductivity,
the real part of the dielectric function, and the loss-function
using these parameters together with the experimental spectra at 4 K.
The damping factors were adjusted to 
the experimental linewidth of the loss-peaks, 
$\gamma_I/2\pi c=\gamma_K/2\pi c=0.5$cm$^{-1}$.

\begin{figure}
 \centerline{\epsfig{figure=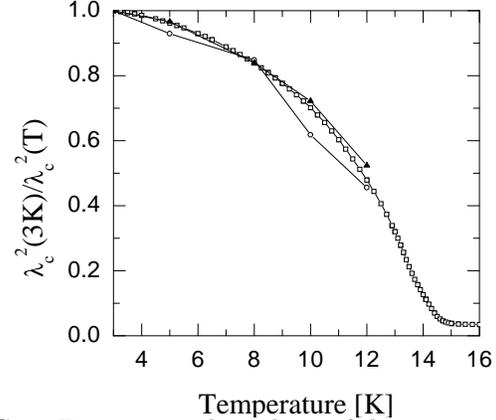,width=7cm,clip=}}
 \caption{Temperature dependence of the $c$-axis penetration depth:
 open squares-measured in cavity perturbation experiment,
 triangles: $\lambda_{c,J}$, open circles: $\lambda_{c,K}$}
 \label{penetrat}
\end{figure}

From the microwave data we have obtained an absolute
measurement\cite{saed} of the $c$-axis penetration depth, which is
45 $\mu$m at 3K. The penetration depth is $\lambda_c =
\lim_{\omega\rightarrow 0}
c/(\omega\mbox{Im}\sqrt{\epsilon(\omega)})$, which in the
framework of Eq.~\ref{eq:difun2} is
$\epsilon^s_{av}\lambda_{c}^2/c^2=\tilde{z}_I\omega_I^{-2}+\tilde{z}_K\omega_K^{-2}$.
With the experimental values for $\tilde{z}_I$, $\omega_I$ and $\omega_K$
determined above, this provides $\lambda_c = 47$ $\mu$m, in
excellent agreement with the microwave results. In
Fig.~\ref{penetrat} the separate temperature dependencies of
$\lambda_{c}$, $\lambda_{c,J}$ and $\lambda_{c,K}$ are given in
the normalized form $\lambda_{c}^2(3~K)/\lambda_{c}^2(T)$;
$\lambda_{c}$ is obtained from a 1~GHz cavity perturbation
measurement, and
 $\lambda_{c,J} = c / (\sqrt{\epsilon_I}\omega_I)$,
 $\lambda_{c,K} = c / (\sqrt{\epsilon_K}\omega_K)$
from the transmission measurement. As seen from
Fig.~\ref{penetrat} both $\lambda_{c,J}$ and $\lambda_{c,K}$
exhibit similar, characteristic, power law temperature dependence
$\lambda_{c}^2(3K)/\lambda_{c}^2(T)=1-T^{\eta}$, $\eta \sim 3$.
This establishes that both resonances are indeed Josephson
plasmons.
%

The intensity of the peaks, represented by the weight factors,
$\tilde{z}_K=0.04$ and $\tilde{z}_I=0.96$,
can be used to estimate the value of the
compressibility of the 2D electron fluids, using the expresions
derived in \cite{lt21app}. The dielectric function
is of the form given in Eq.~\ref{eq:difun2} while
$\tilde{z}_K$ can be calculated from the experimental peak positions 
and linewidths $\omega_K$,$\omega_I$,$\gamma_I$, and $\gamma_K$ using
\begin{equation}
 \begin{array}{ll}
  \tilde{z}_K =& \frac{1}{2}      
    + \frac{(z_K-z_I)(z_Kz_I+2\gamma)}{2(z_Kz_I+2\gamma+4\gamma^2)}
    \sqrt{1-\frac{4(2\gamma)^2\omega_K^2\omega_I^2}
         {(z_Kz_I+2\gamma)(\omega_K^2-\omega_I^2)^2}}\\ 
    &- \frac{2\gamma(z_Kz_I+\gamma)}{z_Kz_I+2\gamma+4\gamma^2}
      \frac{\omega_K^2+\omega_I^2}{\omega_K^2-\omega_I^2}\\                                    
 \end{array}
 \label{tildez}
\end{equation}
Here $\gamma$ is a dimensionless constant which is inversely
proportional to the two-dimensional compressibility of the
electron fluids in the planes, $\gamma = \epsilon^s_{av} /
(4\pi d e^2 Kn^2)$. The constants $z_K$ and $z_I$, defined as 
$z_K=x_K\epsilon^s_{av}/\epsilon^s_{K}$ and
$z_I=\epsilon^s_{av}/\epsilon^s_{J}$ correspond to the weight factors in
the limit of infinite compressibility ($\gamma\rightarrow 0$). 
We can calculate them easily from the known 
volume fractions $x_K$ and $x_I$, (0.5 in the present case) and from 
the ratio between the $T$ -phase and the $T'$ -phase c-axis dielectric constants,
$\epsilon^s_K/\epsilon^s_I$. From infrared spectra of 
$T$ -phase and the $T'$ -phase materials\cite{tajima}, this
ratio is known to be in the interval 
$1 \le \epsilon^s_K/\epsilon^s_I \le 1.7$\cite{tajima}. 
The (numerical) inversion of Eq.~\ref{tildez} then
provides $Kn^2 = 1.12 \pm 0.01 eV^{-1}$ per unit of CuO$_2$.
For La$_{2-x}$Sr$_{x}$CuO$_{4-\delta}$ the linear term in the 
normal state specific heat\cite{loram} ranges from 
0.4 mJ/g.at.K$^2$ (x=0.1) to 1.2 mJ/g.at.K$^2$ (x=0.2).
For fermionic excitations
this would imply that the density of states
varies from 1.2 eV$^{-1}$ (x=0.1) to 5.0 eV$^{-1}$ per unit of CuO$_2$ (x=0.2). 
The value of $Kn^2 = 1.1 eV^{-1}$ should then be taken as an indication, 
that our SmLaSrCuO crystals are underdoped. This is also 
indicated by the T$_c$ of 16 K, and the low values of the JPR-frequencies.

Without the compressibility term, the JPR's would have been 
at $6.1$ cm$^{-1}$ and $8.0$ cm$^{-1}$\cite{lt21app}.
Hence the finite electronic compressibility causes
a large shift of resonance frequency and a large reduction
of the intensity of the higher frequency JPR. This
also explains why the 'bump' has a very
weak intensity in Bi2212\cite{zelezny}.

In conclusion, we have studied microwave and infrared properties
of SmLa$_{0.8}$Sr$_{0.2}$CuO$_{4-\delta}$. We have observed two longitudinal
Josephson plasma resonances, and the
transverse optical plasma mode that arises from the alternation of
two different Josephson couplings between the CuO$_2$
layers. We observe a shift of frequency of the plasma modes, and a
reduction in intensity of the highest mode due to the finite
electronic compressibility corresponding to
$1.1$ eV $^{-1}$ per CuO$_2$ unit.

We gratefully acknowledge Lev N. Bulaevskii for pointing out the
importance of dispersion effects for the optical spectra. One of 
us (DMB) wants to thank the Peter House College for financial
support. This investigation was supported 
by the Netherlands Foundation for Fundamental Research on Matter 
(FOM) with financial aid from the Nederlandse Organisatie voor 
Wetenschappelijk Onderzoek (NWO).


\begin{references}
\bibitem{tony}
A. J. Leggett, Prog. Theor. Phys. {\bf 36}, 901 (1966).
\bibitem{lawrence}
W.E. Lawrence and S. Doniach in {\em Proceedings of the 12th
International Conference on Low Temperature Physics, Kyoto, 1970,}
edited by E. Kanda (Keigaku, Tokyo, 1970).
\bibitem{lt21}
D. van der Marel and A. A. Tsvetkov in {\em Proceedings of the 21t
International Conference on Low Temperature Physics, Prague, 1996,}
edited by S. Danis, V. Gregor, and K. Zaveta; Czech. J. Phys.
{\bf 46}, 3165 (1996).
\bibitem{markus}
M. Gr\"uninger {\em et al.}, Phys. Rev. Lett. {\bf 84}, 1575 (2000).
\bibitem{munzar}
D. Munzar {\em et al.}, Solid State Commun. {\bf 112}, 365 (1999).
\bibitem{zelezny} V. Zelezny {\em et al.},
J. Low Temp. Phys. {\bf 117}, 1019-1024 (1999) .
\bibitem{lt21app}
D. van der Marel, and A. A. Tsvetkov, submitted;
http://xxx.lanl.gov/abs/cond-mat/0102303.
\bibitem{koyama}
T. Koyama, and M. Tachiki, Phys.Rev. B {\bf 54}, 16183 (1996).
\bibitem{tokura}
Y. Tokura {\em et al.}, Phys. rev. B {\bf 40}, 2568 (1989).
\bibitem{shibata}
H. Shibata and T. Yamada, Phys. Rev. Lett. {\bf 81}, 3519 (1998).
\bibitem{andrei}
A.A. Volkov {\em et al.}, Infrared Phys. {\bf 25}, 365 (1985).
\bibitem{tajima}
S. Tajima {\em et al.}, Phys. rev. B {\bf 43}, 10496 (1991).
\bibitem{saed} 
D. Broun and S. Kamal, unpublished.
\bibitem{loram}
J. W. Loram {\em et al.}, Physica C {\bf 235-240} 134 (1994).
\end{references}
\end{document}